# Compressive sensing based Bayesian sparse channel estimation for OFDM communication systems: high performance and low complexity

Guan Gui#, Li Xu*, Lin Shan+, and Fumiyuki Adachi#

#Department of Communications Engineering, Graduate School of Engineering, Tohoku University, Sendai, 980-8579 Japan

*Faculty of Systems Science and Technology, Akita Prefectural University, Akita, 015-0055, Japan

+ Wireless Network Research Institute, National Institute of Information and

Communications Technology (NICT), Yokosuka, 239-0847, Japan

Email: gui@mobile.ecei.tohoku.ac.jp

**Abstract**

In orthogonal frequency division modulation (OFDM) communication systems, channel state information (CSI) is required at receiver due to the fact that frequency-selective fading channel leads to disgusting inter-symbol interference (ISI) over data transmission. Broadband channel model is often described by very few dominant channel taps and they can be probed by compressive sensing based sparse channel estimation (SCE) methods, e.g., orthogonal matching pursuit algorithm, which can take the advantage of sparse structure effectively in the channel as for prior information. *However, these developed methods are vulnerable to both noise interference and column coherence of training signal matrix*. In other words, the primary objective of these conventional methods is to catch the dominant channel taps without a report of posterior channel uncertainty. *To improve the estimation performance, we proposed a compressive sensing based Bayesian sparse channel estimation (BSCE) method which can not only exploit the channel sparsity but also mitigate the unexpected channel uncertainty without scarifying any computational complexity*. The proposed method can reveal potential ambiguity among multiple channel estimators that are ambiguous due to observation noise or correlation interference among columns in the training matrix. Computer simulations show that propose method can improve the estimation performance when comparing with conventional SCE methods.

## 1. Introduction

In broadband wireless communication systems using orthogonal frequency division modulation (OFDM), frequency-selective fading is incurred by the reflection, diffraction and scattering of the transmitted signals due to the buildings, large moving vehicles, mountains, etc. Such fading phenomenon distorts received signals and poses critical challenges in the design of communication systems for high-rate and high-mobility wireless communication applications. Hence, accurate channel estimation becomes a fundamental problem of such communication systems. In last several years, various linear estimation methods have been proposed based on the assumption of rich multipath channel model. However, recently, a lot of physical channel measurements verified the channel taps exhibit sparse distribution [1]–[3] due to the broadband signal transmission. A typical example of sparse multipath channel is shown in Fig. 1 where the length is 100 while the number nonzero taps is 5 only. Note that different broadband transmission may incur different channel structures in wireless communications systems as shown in Tab. 1.

To improve the estimation performance, extra sparse structure information can be exploited as prior information. Thanks to the development of compressive sensing [4], [5], many sparse channel estimation (CCE) methods have been proposed for exploiting the channel sparsity. In [6], orthogonal matching pursuit (OMP) algorithm with application to sparse multipath channel estimation in the OFDM systems. In [7][8], sparse channel estimation methods have been proposed using compressive sampling matching pursuit (CoSaMP) algorithm [9] in frequency-selective and doubly-selective channel fading communication systems. In [10], to further reduce the computational complexity, sparse channel estimation using smooth $\ell_0$-norm (SL0) algorithm [11] has been proposed. Compared to traditional linear methods, sparse channel estimation methods have two obvious advantages: spectral efficiency and lower performance bound. For one thing, improve the spectral efficiency by utilizing less training sequence can achieve the same estimation performance as linear methods. For another, obtain the lower performance bound by exploiting channel sparsity due to the fact less active channel freedom of degree is acquired [12].

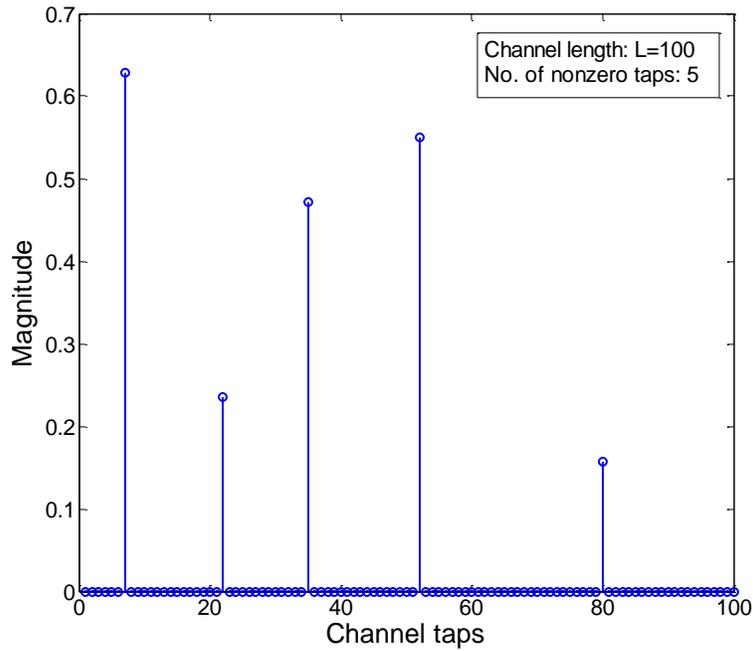

Fig. 1. A typical example of sparse multipath channel.

Conventional sparse channel estimation methods have a cardinal objective that try to probe the dominant channel taps as much as accurate, while neglect the posterior information report from additive noise received signal. These proposed channel estimation methods are termed as model selection or basis selection. Unfortunately, their estimation performances are often degraded due to the neglecting channel model uncertainty [13][14]. To mitigate the unexpected model uncertainty, Bayesian compressive sensing (BCS) [15] and an slight improved Bayesian compressive sensing using laplace priors (BCS-LAP) [16] could be adopted for estimating sparse channel. The estimation performance could be improved effectively but at the cost of high computational complexity when comparing with existing simple algorithms (e.g., OMP [6] and SL0 [11]). Hence, it is impractical to employ this algorithm in real communication systems.

Table 1. Channel structures in different mobile communication systems.

| Generations of mobile communication systems [17] | 2G Cellular (IS-95) | 3G Cellular (WCDMA) | 4G/5G Cellular (LTE-Advanced~) |
|---|---|---|---|
| Transmission bandwidth | 1.23MHz | 10MHz | 20MHz~100MHz |
| Time delay-spread (assume) | $0.5\ \mu s$ | $0.5\ \mu s$ | $0.5\ \mu s$ |
| Sampling channel length | 1 | 10 | 20~100 |
| Number of nonzero taps | 1 | 4 | 2~10 |
| Channel structure model | dense | approximate sparse | sparse |

Unlike these aforementioned methods, in this paper, we propose an improved Bayesian sparse channel estimation (BSCE) method while its computational complexity is comparable with OMP and SL0. Our proposed Bayesian channel method can be divided into two steps: position detection of dominant channel taps and channel estimation using minimum mean square error (MMSE). In general, our proposed Bayesian estimation method provides model uncertainty which reveals uncertainty among multiple possible position sets of dominant channel taps that are ambiguous due to observation noise or correlation among columns in the training matrix. Furthermore, the complexity of the propose method is relative lower due to its smaller search space when compare to conventional methods. Simulation results are given to verify two folds: performance and complexity. Note that estimation performance is evaluated by two metrics: mean-square-error (MSE) and bit-error rate (BER), while computational complexity is measured coarsely by CPU time of computer.

The remainder of this paper is organized as follows. An simple OFDM system model is described and problem formulation is given in Section 2. In section 3, the BSCE method is proposed in OFDM systems. Computer simulation results are given in Section 4 in order to evaluate and compare performance of the BSCE method with conventional methods. Finally, we conclude the paper in Section 5.

*Notation*: Throughout the paper, matrices and vectors are represented by boldface upper case letters (i.e., $\boldsymbol{X}$) and boldface lower case letters (i.e., $\boldsymbol{x}$), respectively; the superscripts $(\cdot)^T$, $(\cdot)^H$, $(\cdot)^{-1}$ and $\text{diag}(\cdot)$ denote the transpose, the Hermitian transpose, the inverse and diagonal operators, respectively; $E\{\cdot\}$ denotes the expectation operator; $\|\boldsymbol{h}\|_0$ is the $\ell_0$-norm operator that counts the number of nonzero taps in $\boldsymbol{h}$ and $\|\boldsymbol{h}\|_p$ stands for the $\ell_p$-norm operator which is computed by $\|\boldsymbol{h}\|_p = (\Sigma_l |h_l|^p)^{1/p}$, where $p \in \{1, 2\}$ is considered in this paper.

## 2. System model and problem formulation

Consider a frequency-selective multipath channel whose impulse response is given by

$$\boldsymbol{h} = \sum_{l=0}^{L-1} h_l \delta(\tau - \tau_l), \tag{1}$$

where $L$ is the number of multipaths, and $h_l$ and $\tau_l$ are the (complex) channel gain and the delay spread, respectively, of path $l$ at time $t$. Hence, the $L$-length discrete channel vector can be written as $\boldsymbol{h} = [h_0, h_1, \cdots, h_{L-1}]^T$. Let the OFDM system uses size-$N$ discrete Fourier transform (DFT) and its number of pilot subcarriers is $N_p$. To avoid inter-symbol interference (ISI), we assume that the length $N_g$ of the zero-padding cyclic prefix (CP) in the OFDM symbols is larger than maximum delay spread $\tau_{\max}$, where $\tau_{\max} \geq \tau_l$, $l = 0, 1, \cdots, L-1$. Suppose that $\bar{X}(i)$ denote $i$-th subcarrier in an OFDM symbol, where $i = 0, 1, \cdots, N-1$. If the coherence time of the channel is much larger than the OFDM symbol duration $T$, then the channel can be considered quasi-static over an OFDM symbol. Let $\bar{\boldsymbol{y}}$ be the vector of received signal samples in one OFDM symbol after DFT, then

$$\bar{\boldsymbol{y}} = \bar{\boldsymbol{X}}\bar{\boldsymbol{h}} + \bar{\boldsymbol{z}} = \bar{\boldsymbol{X}}\boldsymbol{F}\boldsymbol{h} + \bar{\boldsymbol{z}} = \boldsymbol{X}\boldsymbol{h} + \boldsymbol{z}, \tag{2}$$

where $\bar{\boldsymbol{X}} = \text{diag}\{X(0), X(1), \ldots, X(N-1)\}$ denotes diagonal subcarrier matrix, $\bar{\boldsymbol{h}}$ is the channel frequency response (CFR) in frequency-domain, $\bar{\boldsymbol{z}}$ is assumed to be additive white Gaussian noise

(AWGN) with variance $\sigma^2$. $F$ is an $N \times L$ partial DFT matrix with its $k$-th row which is easily given by $1/\sqrt{N}\left[0, e^{-j2\pi k/N}, \cdots, e^{-j2\pi k(L-1)/N}\right]$ and $X = \bar{X}F = [x_0, \cdots, x_l, \cdots, x_{L-1}]$ denotes an $N \times L$ equivalent time-domain signal matrix. In addition, $h = [h_0, h_1, \ldots, h_{L-1}]^T$ denotes a $L \times 1$ time-domain channel vector. Since $\bar{h} = Fh$, hence, the frequency-domain channel impulse response $\bar{h}$ lies in the time-delay spread domain.

Assuming a binary random vector $g = [g_0, g_1, \cdots, g_{L-1}]^T$ denote an taps' position indicator of sparse channel vector $h$ which is generated from a Gaussian mixture density (GMD) function as

$$\{h|g\} \sim \mathcal{CN}(0, R(g)), \tag{3}$$

where the covariance matrix $R(g)$ is determined by position indicator $g$. For a better understanding, we take $R(g)$ to be diagonal element with $[R(g)]_{ll} = \sigma_l^2 = \sigma_1^2$ for $l = 0, 1, \cdots, L-1$, implying that $\{h_l|g_l\}_{l=0}^{L-1}$ are independent with Gaussian distribution $\{h_l|g_l\} \sim \mathcal{CN}(0, \sigma_1^2)$. Assume that the position indices $\{g_l\}_{l=0}^{L-1}$ are satisfied Bernoulli distribution with probability $p_{1,l}$, then the probability of nonzero and zero channel taps of channel vector $h$ can be written as

$$\begin{cases} h_l \neq 0 \Leftarrow \Pr\{g_l = 1\} = p_{1,l} \\ h_l = 0 \Leftarrow \Pr\{g_l = 0\} = 1 - p_{1,l} \end{cases} \tag{4}$$

for $l = 0, 1, \cdots, L-1$. According to (4), one can easily find $\|h\|_0 = \|g\|_1$. In real communication systems, broadband channels are often described by sparse models [18], [19]. Hence, we choose $\sigma_0^2 = \text{var}\{h_0|g_0\} = 0$ and $p_1 = \sum_{l=0}^{L-1} p_{1,l} \ll 1$, which equivalents to that the channel vector $h$ has relatively few dominant channel taps. In other words, sparseness of channel vector $h$ depends on the probability $p_1$ as shown in Fig. 2. Smaller probability $p_1$ implies sparser channel vector $h$ and vice versa.

The research objective of this paper is to estimate the sparse channel vector $h$ using received signal vector $y$ and training signal matrix $X$. Hence, the system model can be assumed satisfying distribution as

$$\begin{bmatrix} y \\ h \end{bmatrix} \Big| g \sim \mathcal{CN} \left( 0, \begin{bmatrix} C(g) & XR(g) \\ R(g)X^T & R(g) \end{bmatrix} \right) = \mathcal{CN} \left( 0, \begin{bmatrix} C(g) & \sigma_1^2 X I_L \\ \sigma_1^2 I_L X^T & \sigma_1^2 I_L \end{bmatrix} \right), \quad (5)$$

where $C(g) := XR(g)X^T + \sigma_n^2 I_N = \sigma_1^2 X I_L X^T + \sigma^2 I_N$ is the covariance matrix of $\{y|g\}$. That is $\{y|g\} \sim \mathcal{CN}\left(0, \sigma_1^2 X I_L X^T + \sigma^2 I_L\right)$.

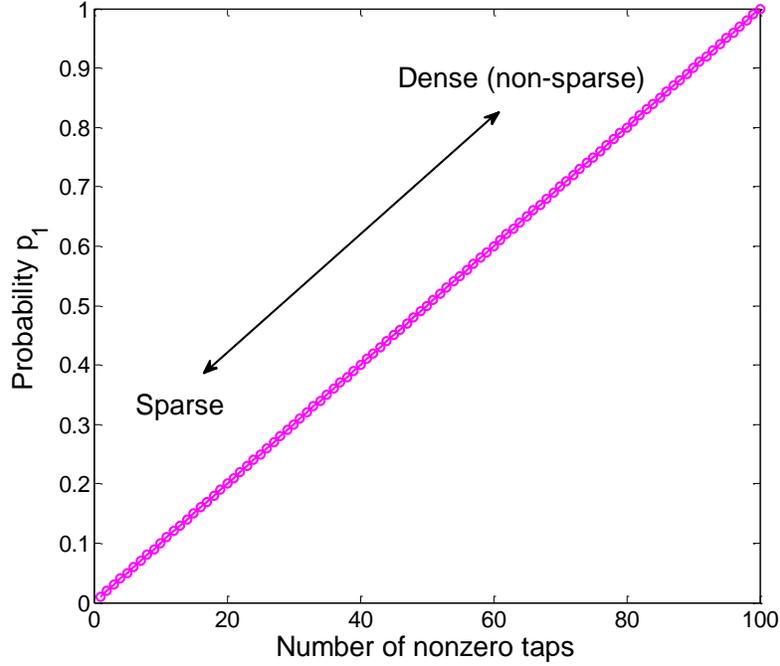

Fig. 2. Sparseness of channel vector $h$ depends on the probability $p_1$.

## 3. Compressive sensing based Bayesian sparse channel estimation

In this section, compressive sensing based Bayesian sparse channel estimation is proposed in two steps: 1) *detect the position set of dominant channel taps* and 2) *then estimate sparse channel $\tilde{h}$ using MMSE algorithm*. Obviously, how to find the dominant channel taps' position is a key technique with low-complexity Bayesian method for estimating sparse channels.

*3.1. Position detection on dominant channel taps*

According to the well-known Bayesian rules, the posterior of position indicator $g$ can be written as

$$P(g|y) = \frac{P(y|g)P(g)}{\sum_{g' \in G} P(y|g')P(g')}, \qquad (6)$$

where $G = \{0,1\}^L$ denote all of possible position index sets of channel taps as shown in Fig. 3. Eq. (6) implies that estimating $\{P(g|y)\}_{g \in G}$ reduces to estimating $\{P(y|g)P(g)\}_{g \in G}$. Due to the extremely computational complexity in (6), the huge size of $G$ makes it impractical to compute $P(g|y)$ or $\{P(y|g')P(g')\}$ for all $g' \in G$ in the case of high-dimensional broadband channels. By considering sparse structure in channels, only posteriors of dominant taps' position are needed for sparse channel estimation. Assuming the set $G_*$ responsible for position indicator of dominant channel taps, then the search space in $G_*$ rather than $G$ can be quite small and therefore practical to compute. Hence, the posteriors of dominant channel taps can be approximated by

$$P(g|y) \approx \frac{P(y|g)P(g)}{\sum_{g' \in G_*} P(y|g')P(g')}, \qquad (7)$$

for dominant channel set $G_*$. Hence, exploiting the dominant channel taps set $G_*$ reduces to the search for $g \in G_*$ which only computes the dominant values of $P(y|g)P(g)$ in (7). First of all, the probability density function (PDF) $P(y|g)$ for position indicator $g \in G_*$ can be written as

$$P(y|g) = \frac{1}{\sqrt{(2\pi)^L \det(C(g))}} \exp\left(-\frac{1}{2} y^T C^{-1}(g) y\right). \qquad (8)$$

By transformed it in log-domain for convenience, then the position indicator (PI) $\text{PI}(g,y)$ can be given by

$$\begin{aligned}\text{PI}(g,y) &\triangleq \ln P(y|g)P(g) = \ln P(y|g) + \ln P(g) \\ &= \ln P(y|g) + \|g\|_0 \ln p_1 + (L - \|g\|_0)\ln(1-p_1) \\ &= -\frac{L}{2}\ln 2\pi - \frac{1}{2}\ln \det(C(g)) - \frac{1}{2} y^T C^{-1}(g) y \\ &\quad + \|g\|_0 \ln \frac{p_1}{1-p_1} + L\ln(1-p_1), \end{aligned} \qquad (9)$$

which is a metric of position indicator $g$. According to $\text{PI}(g,y)$ in (9), one can easily find that the position indicator depends on received signal, channel length, position indicator and probability of nonzero taps. Due to the positive exponent relationship $P(g|y) = e^{\text{PI}(g,y)}$, $\text{PI}(g,y)$ in (9) can also be considered as a measure function of $P(g|y)$ on dominant channel taps. However, it is still unfeasible to get the position information of channel in practical system without considering channel estimation. According to [20], the mathematical expectation of $\text{PI}(g,y)$ can be given by

$$E\{\text{PI}(g,y)\} = 2N + Lp_1(1-p_1)\left(\ln\left[\left(\sigma_1^2/\sigma^2 + 1\right)(1-p_1)/p_1\right]\right)^2. \qquad (10)$$

For a given pair $\{g',y\}$, $\text{PI}(g',y)$ can be used to compare the mean $E\{\text{PI}(g',y)\}$ and standard deviation $\sqrt{\text{var}\{\text{PI}(g',y)\}}$ in order to get a rough evaluation of $(g',y)$ which whether or not has a dominant probability.

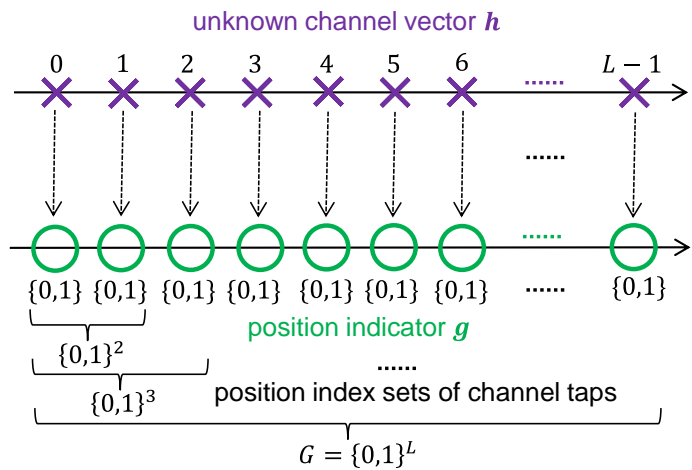

Fig. 3. Graph illustration for all of possible position index set of channel taps.

To reduce the search space in position set, we resort to an efficient method [13] to determine $G_*$ as follows. The basic idea is that the position set $g$ of unknown channel yielding the dominant values of $P(g|y)$ is equivalent to the high probability of $\text{PI}(g,y)$. The search starts with $g = 0$ and the initial position set is set as $G^{(0)}$. If we change each element in $g$ and then it yields $L$ position indicators. Consider all of position indicators in a set and refer it to $G^{(1)}$. The metrics $\text{PI}(g,y)$ for the $L$ PI vectors in $G^{(1)}$ are then computed by (9), and elements of $G^{(1)}$ with the $D$ largest value of the dominant channel tap are collected in $G_*^{(1)}$. For each possible dominant taps' set in $G_*^{(1)}$, all positions of a second nonzero tap are then considered, yielding $\sum_{i=1}^{D}(L-i) = LD - D(D+1)/2$ unique binary vectors to store in $G^{(2)}$. The $\text{PI}(g,y)$ for all vector in $G^{(2)}$ are then computed, and the elements of $G^{(2)}$ with the $D$ largest value are collected in $G_*^{(2)}$. Then for each candidate vector in $G_*^{(2)}$, all possibilities of a third dominant channel tap are considered, and those with the $D$ largest channel taps are stored in $G_*^{(3)}$. The process continues until $G_*^{(S)}$ is computed, where $S$ can be chosen to make $\Pr(\|h\|_0 > S)$ sufficiently small to exploit all of channel sparsity. Note that $G_*^{(S)}$ constitutes the algorithm's final estimate of $G_*$ and later we denote $\hat{G}_*$ as the final estimate. For bettering understanding the PI update of dominant channel taps, an intuitive example is given in Fig. 4, where the length of position indicator $g$ is set as $L = 5$; number of largest value of PI is choose as $D = 1$ and maximum number of nonzero taps is set as $S = 3$.

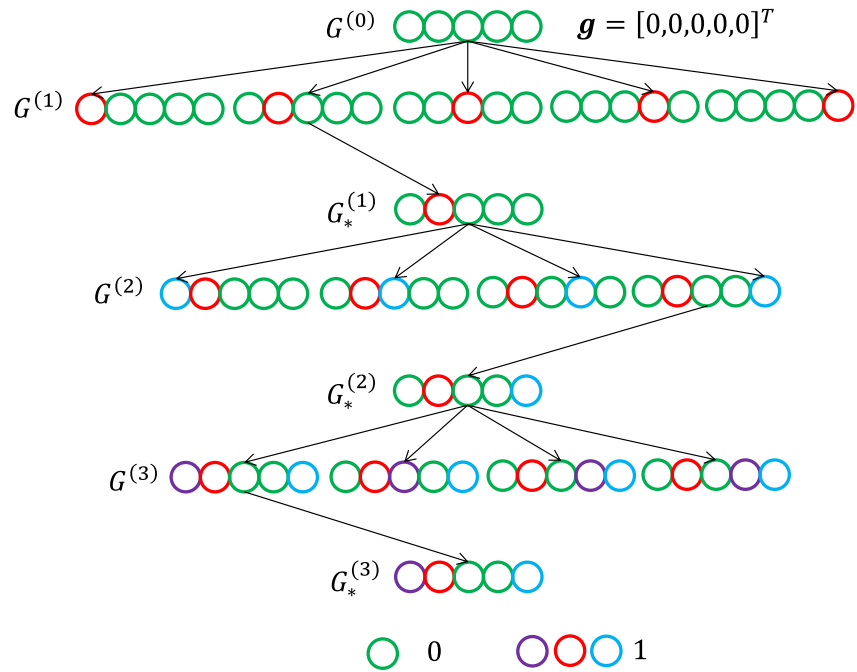

Figure 4. An intuitive example of position set selection on dominant channel taps, where the green circle denotes zero while the other color circles denote one.

For use with the aforementioned Bayesian matching pursuit (BMP) algorithm, we consider a fast metric update which computes the change in $\text{PI}(\cdot)$ that results from the activation of a position of nonzero tap. More precisely, if we denote by $g_l$ the vector identical to $g$ except for the $l$-th coefficient, which is active in $g_l$ but inactive in $g$ (i.e., $[g_l]_l = 1$ and $[g]_l = 0$), then it is defined as

$$d_l(g) \triangleq \text{PI}(g_l, y) - \text{PI}(g, y), \tag{11}$$

to track the change of active positions. Note that the $\text{PI}(g, y)$ at the initial step is set as

$$\text{PI}(0, y) = -\frac{L}{2}\ln 2\pi - \frac{N}{2}\ln \sigma_1^2 - \frac{1}{2\sigma^2}\|y\|_2^2 + L\ln(1 - p_1), \tag{12}$$

via (9) and the fact that $C(0) = \sigma_1^2 I_L$. To obtain the fast PI update, we start with the property that, for any $l$ and $g$,

$$C(g_l) = C(g) + \sigma_1^2 x_l x_l^T, \tag{13}$$

for which the matrix inversion lemma implies

$$C^{-1}(g_l) = C^{-1}(g) - \sigma_1^2 \beta_l b_l b_l^T, \tag{14}$$

$$C^{-1}(g) = \frac{1}{\sigma^2}I_N - \sigma_1^2 \sum_{i=1}^{P} \beta^{(i)} b^{(i)} (b^{(i)})^T \tag{15}$$

$$b_l \triangleq C^{-1}(g)x_l = \frac{1}{\sigma^2}x_n - \sigma_1^2 \sum_{i=1}^{P} \beta^{(i)} b^{(i)} (b^{(i)})^T x_l \tag{16}$$

where $b_l := C^{-1}(g)x_l$ and $\beta_l := (1 + \sigma_1^2 x_l^T b_l)^{-1}$. Notice that the cost of computing $\beta_l$ in (14) is $\mathcal{O}(LN^2)$ if standard matrix multiplication is used [13]. According to previus analysis, we can get

$$\begin{aligned} y^T C^{-1}(g_l) y &= y^T \left( C^{-1}(g) - \sigma_1^2 \beta_l b_l b_l^T \right) y \\ &= y^T C^{-1}(g) y - \sigma_1^2 \beta_l \left( y^T b_l \right)^2, \end{aligned} \tag{17}$$

$$\begin{aligned} \ln\det\left(C(g_l)\right) &= \ln\det\left(C(g) + \sigma_1^2 x_l x_l^T\right) \\ &= \ln\left[(1 + \sigma_1^2 x_l^T C^{-1}(g) x_l) \det\left(C(g)\right)\right] \\ &= \ln\det\left(C(g)\right) - \ln\beta_l, \end{aligned} \tag{18}$$

$$\|g_l\|_0 \ln\frac{p_1}{1-p_1} = \left(\|g\|_0 + 1\right)\ln\frac{p_1}{1-p_1} = \|g\|_0 \ln\frac{p_1}{1-p_1} + \ln\frac{p_1}{1-p_1}, \tag{19}$$

which, combined with (5), yield

$$\mathrm{PI}(g_l, y) = \mathrm{PI}(g, y) + \underbrace{\frac{1}{2}\ln \beta_l + \frac{\sigma_1^2}{2}\beta_l \left(y^T b_l\right)^2 + \ln \frac{p_1}{1-p_1}}_{d_l(g)}. \tag{20}$$

In summary, $d_l(g)$ in (18) quantifies the change in $\mathrm{PI}(\cdot)$ due to the activation of the $l$-th position of $g$.

Please note that the cost of computing $\{\beta_l\}_{l=0}^{L-1}$ via $b_l := C^{-1}(g)x_l$ and $\beta_l := (1 + \sigma_1^2 x_l^T b_l)^{-1}$ is $\mathcal{O}(LN^2)$, if standard matrix multiplication is used. As we describe, the complexity of this operation can be made linear in $N$ by exploiting the structure of $C^{-1}(g)$. Say that $t = [t_1, t_2, \cdots, t_p]^T$ contains the indices of active elements in $g$. Then from (14), we can get

$$C^{-1}(g) = \frac{1}{\sigma^2}I_N - \sigma_1^2 \sum_{i=1}^{p} \beta^{(i)} b^{(i)} \underbrace{b^{(i)T} x_l}_{:=c_l^{(i)}} \tag{21}$$

when activating the $l$-th position in $g$. The key observation is that the coefficients $\{c_l^{(i)}\}_{l=0}^{L-1}$ need only be computed once, i.e., when index $t_i$ is active. Furthermore, $\{c_l^{(i)}\}_{l=0}^{L-1}$ only need to be computed for surviving indices $t_i$. According to previous analysis in (20), the number of multiplications required by the algorithm is $\mathcal{O}(LNPD)$ [13]. Moreover, the complexity of the proposed algorithm could be reduced if the smaller $D$ is adopted.

*3.2. MMSE for estimating values of dominant channel taps*

By utilizing the dominant taps' posteriors, the sparse channel can be estimated readily by MMSE algorithm as

$$\begin{aligned}\tilde{h} &= E\{h|y\} \\ &= \sum_{g \in G} P(g|y) E\{h|y, g\} \\ &\approx \sum_{g \in G_*} P(g|y) E\{h|y, g\}.\end{aligned} \tag{22}$$

According to above introduction, compressive sensing based Bayesian sparse channel estimation could be implement by (20)-(22) with high estimation performance and low complexity.

## 4. Computer Simulations

In this section, the proposed BSCE estimator using 1000 independent Monte-Carlo runs for averaging. The length of channel vector $h$ is set as $N = 100$. Values of dominant channel taps follow Gaussian distribution and their positions are randomly allocated within the length of $h$ which is subjected to $E\{\|h\|_2^2 = 1\}$. The received signal-to-noise ratio (SNR) is defined as $10\log(E_b/\sigma_n^2)$, where $E_b = 1$.

Table 2. Simulation parameters.

| | | |
|---|---|---|
| Transmitter | Data modulation | BPSK |
| | No. of subcarrier | $N_d = 256$ |
| | No. of pilot symbol | $N \in \{20,30,40\}$ |
| | Length of CP | $N_g = 16$ |
| | Pilot sequence | Random Gaussian sequence |
| Channel model | Fading | Frequency-selective block |
| | No. of channel taps | $L = 100$ |
| | Prob. of nonzero taps | $p \in \{0.1, 0.2\}$ |
| | Power delay profile | Random Gaussian |
| Receiver | Channel estimation | BSCE |
| | Data detection | Zero forcing |

The proposed method is compared to five conventional sparse channel estimation methods using algorithms: OMP [21], CoSaMP [9], BCS [15], BCS-LAP [16] and SL0 [22]. It was worth noting that these simulation parameters were chosen in accordance with detailed communication environment in this paper. The stopping error criteria threshold is set as $10^{-4}$ for all algorithms in Monte Carlo computer simulations. The initial noise variance for BSC and BSC-LAP is set as $\text{var}(y)/10$, where $\text{var}(y) = (1/(N-1)\sum_{n=1}^{N}(y_n - \hat{y}))^{1/2}$ denotes standard derivation and $\hat{y} = 1/N \sum_{n=1}^{N} y_i$. In addition, the Laplace prior for BCS-LAP is computed automatically which was suggested in [16]. The parameters of FBMP algorithm were initialized *as* $\lambda_1 = 0.01$, $\mu_1 = 0$, $\sigma^2 = 0.05$, and $\sigma_1^2 = 2$. Computer simulation parameters are listed in Tab. 2.

*A. MSE versus SNR*

The estimation performance is evaluated by average mean square error (MSE) standard which is

defined as

$$MSE\{\tilde{\mathbf{h}}\} = E\left\|\mathbf{h} - \tilde{\mathbf{h}}\right\|_2^2, \qquad (23)$$

where $E\{\cdot\}$ denotes expectation operator, $\mathbf{h}$ and $\hat{\mathbf{h}}$ are the actual channel vector and its channel estimator, respectively. In Figs. 5~8, we compare the average MSE performance of the proposed channel estimator with traditional sparse channel estimators with respect to different channel sparseness, $p_1 = 0.1$ and $p_1 = 0.2$. As the two figures show, our proposed method can achieve better estimation performance than conventional methods. The lower bound is given by least square (LS) method (oracle) which utilized the channel position information. In this figure, it is easy found that proposed method obtained lower MSE performance than conventional methods. In other words, if the proposed estimator is applied in data detection, smaller BER performance can be achieved when comparing with conventional methods.

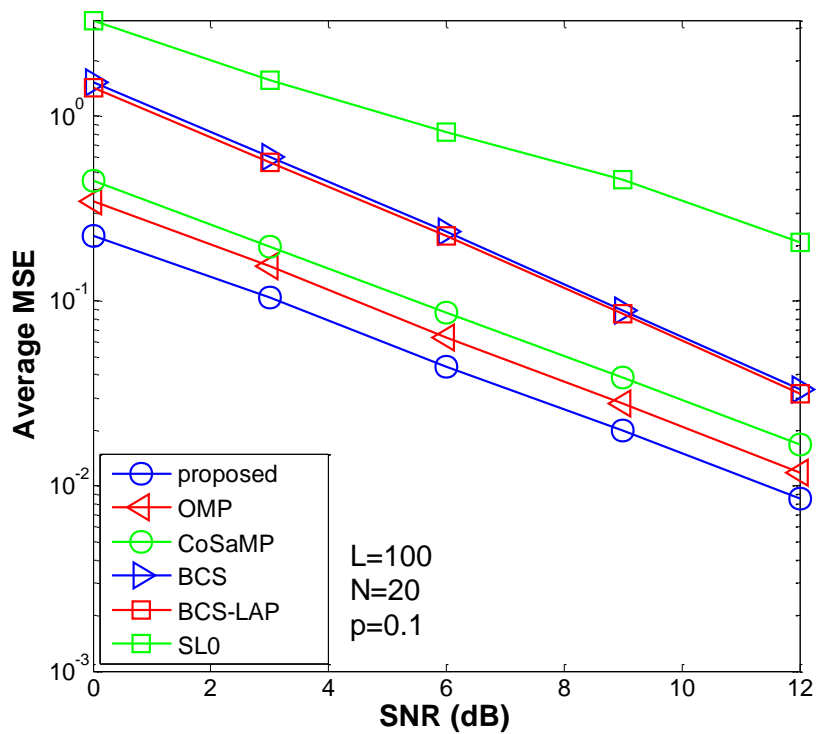

Fig. 5. Average MSE performance verses SNR when $p_1 = 0.1$ and $N = 20$.

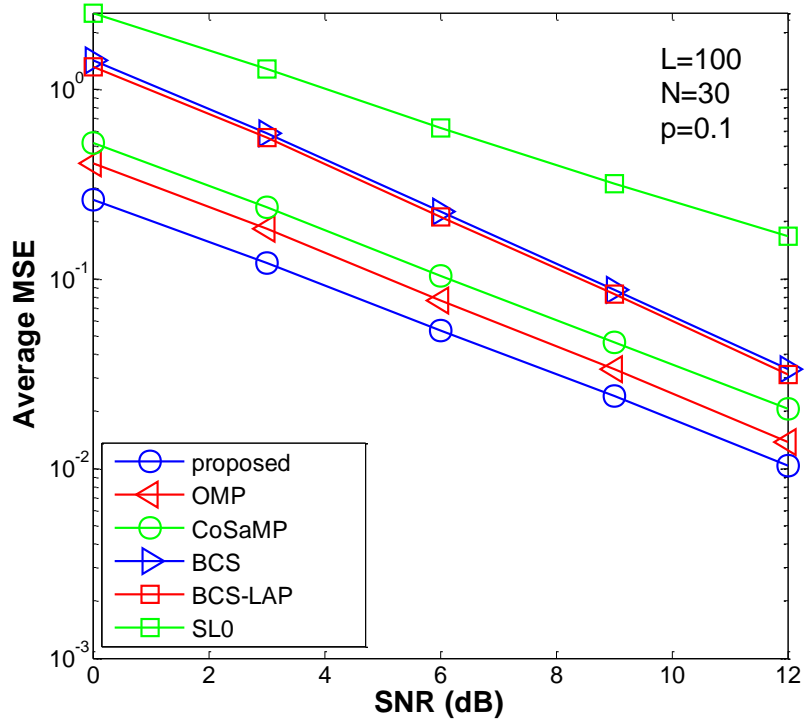

Fig. 6. Average MSE performance verses SNR when $p_1 = 0.1$ and $N = 30$.

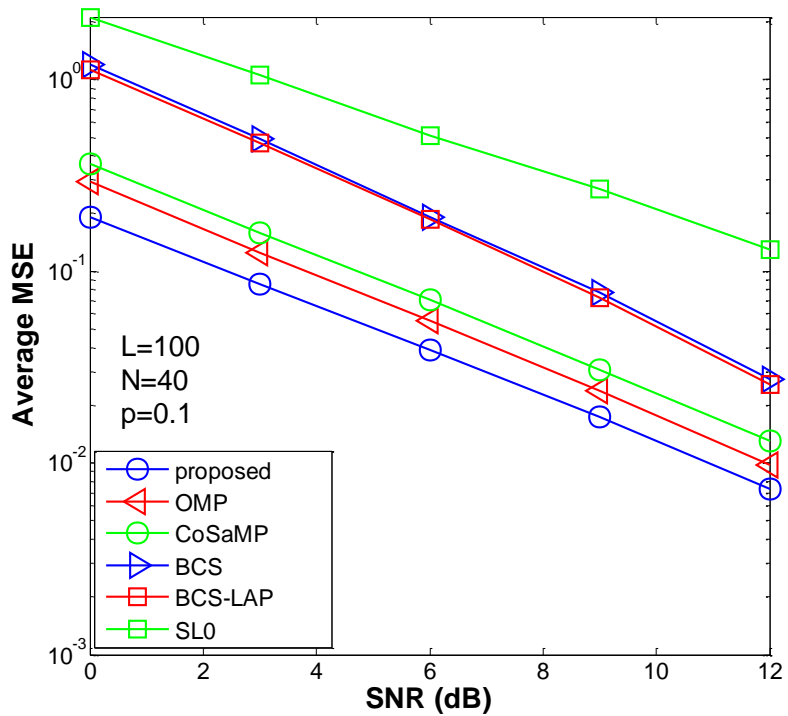

Fig. 7. Average MSE performance verses SNR when $p_1 = 0.1$ and $N = 40$.

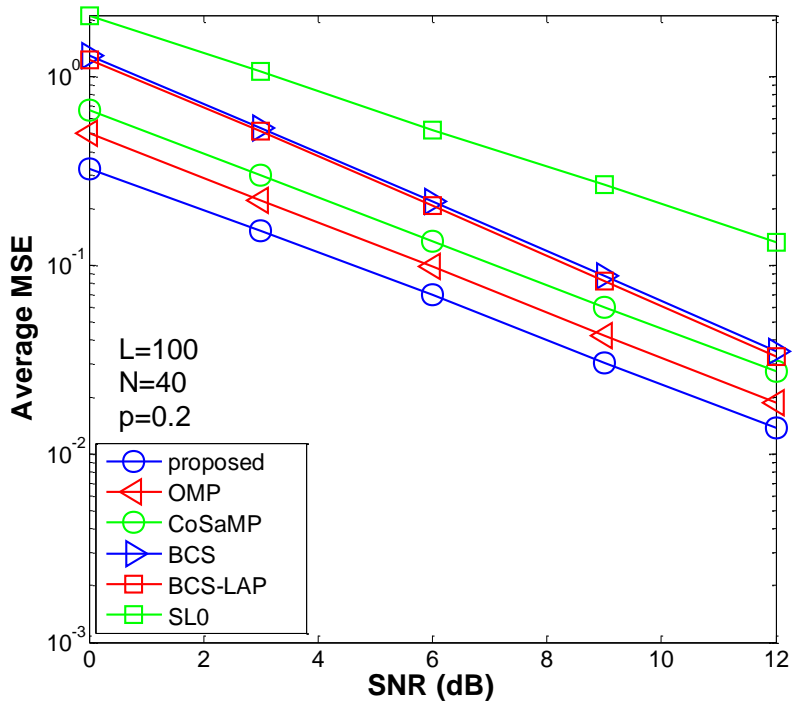

Fig. 8. Average MSE performance verses SNR when $p_1 = 0.2$ and $N = 40$.

## B. BER versus SNR

By using above channel estimators, signal transmission performances are evaluated as shown in Figs. 9~12. From the two figure, average BER performance curves are depicted with respect to SNR for binary phase shift keying (BPSK) data. We can see that the BER performance of the proposed method is more close to lower bound which is given by ideal channel estimator whose nonzero taps' positions are known. Here, only low signal modulation was considered for BER evaluation. It is very easy to predict that our proposed method could improve BER performance in case of high signal modulation.

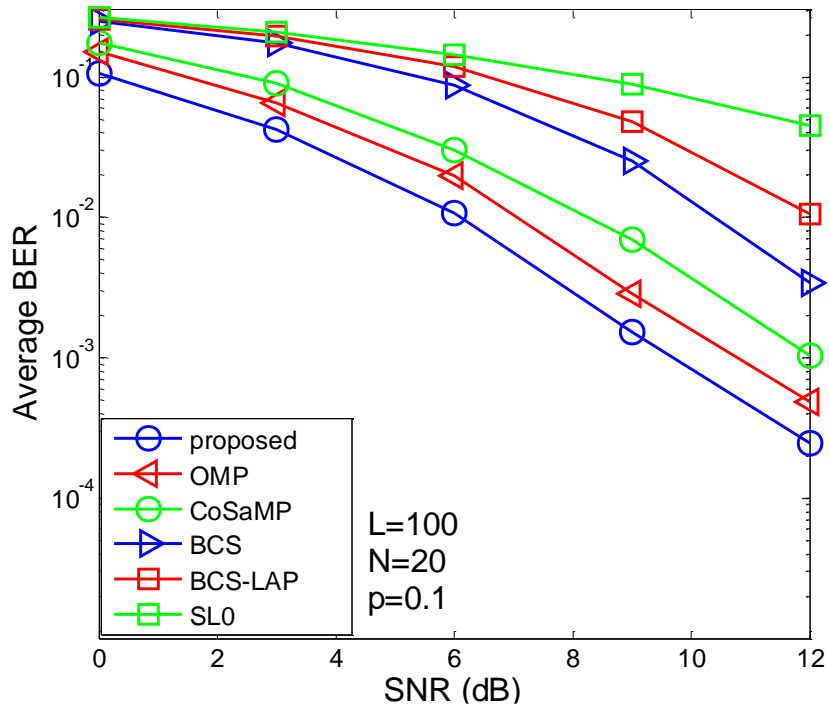

Fig. 9. Average BER performance verses SNR when $p_1 = 0.1$ and $N = 20$.

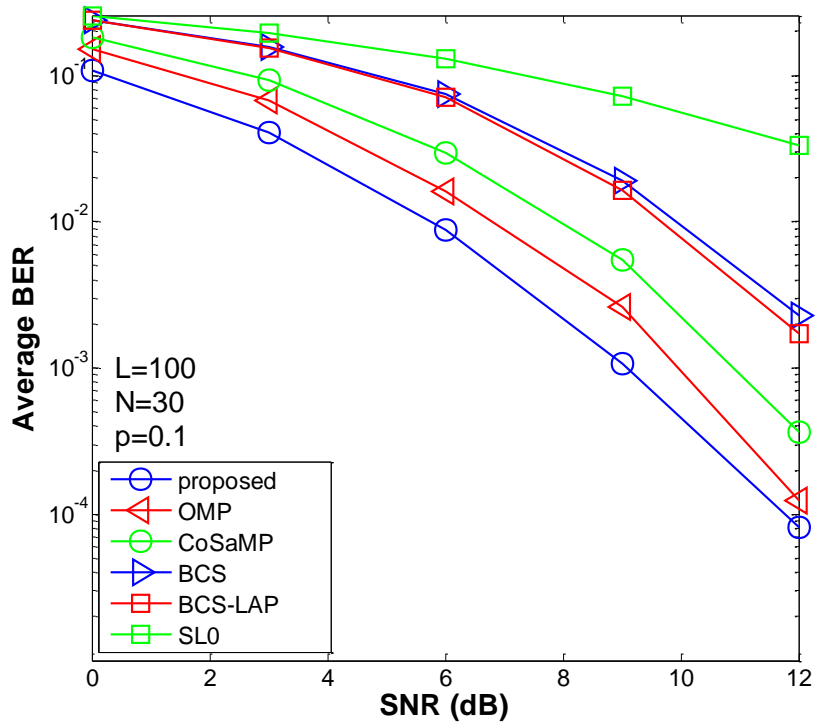

Fig. 10. Average BER performance verses SNR when $p_1 = 0.1$ and $N = 30$.

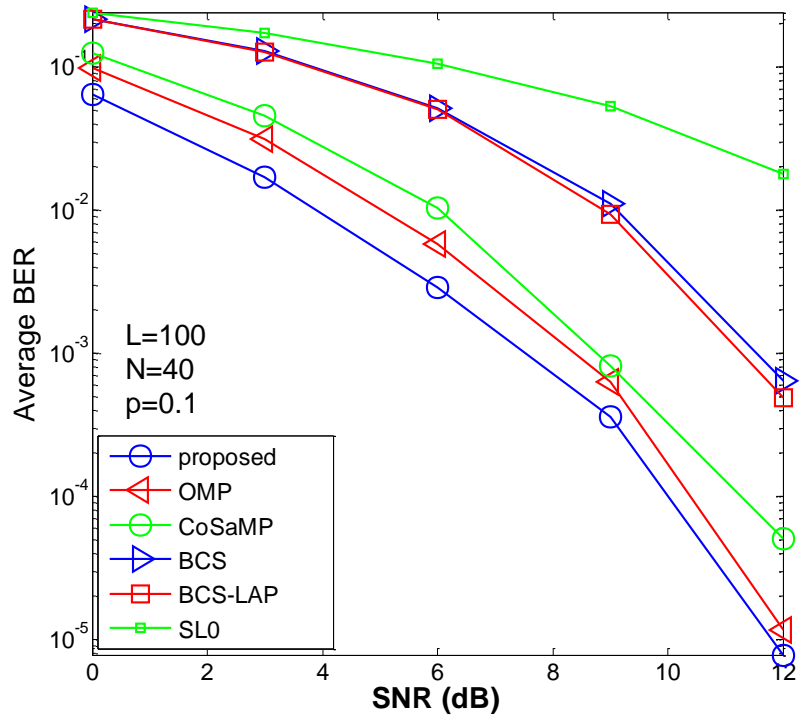

Fig. 11. Average BER performance verses SNR when $p_1 = 0.1$ and $N = 40$.

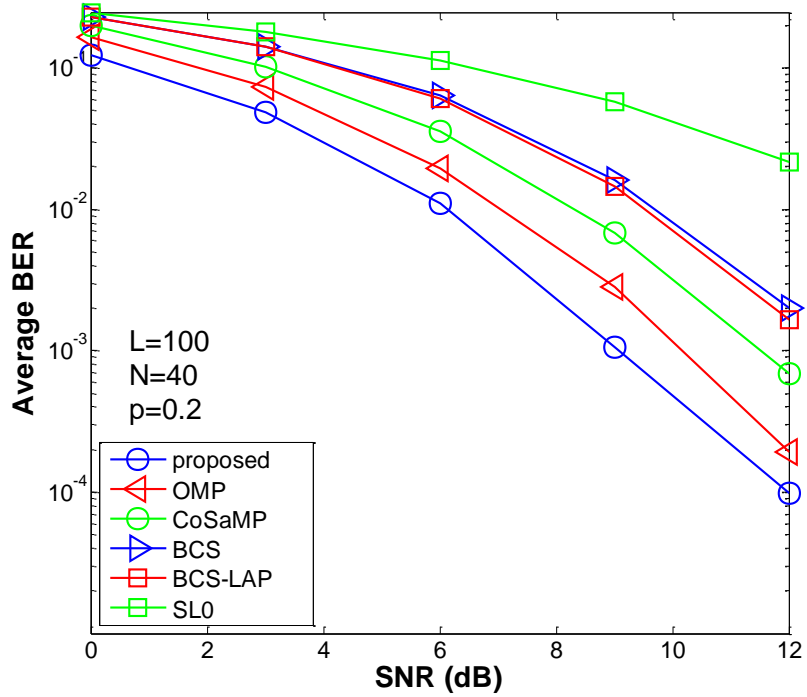

Fig. 12. Average BER performance verses SNR when $p_1 = 0.2$ and $N = 40$.

## C. Complexity evaulation

To compare the computational complexity of the proposed method with other methods, CPU time is adopted for evaluation standard as shown in Figs. 13~16. It is worth mentioning that although the CPU time is not an exact measure of complexity, it can give us a rough estimation of computational complexity. Our simulations are performed in MATLAB 2012 environment using a 2.90GHz Intel i7 processor with 8GB of memory and under Microsoft Windows 8 (64 bit) operating system. For comprehensive comparing this proposed method with other methods in different length of training signal and different channel sparsity, we simulate their comparison results in Figs. 13-16. As the four figures shown, the complexity of the proposed method is close to OMP and SL0-based methods and lower than CoSaMP, BCS and BCS-LAP based methods. It is well known that the complexity of OMP and SL0 is very low on sparse channel estimation [10][23]. Hence, comparing with traditional methods, our proposed method can achieve better estimation performance and low complexity.

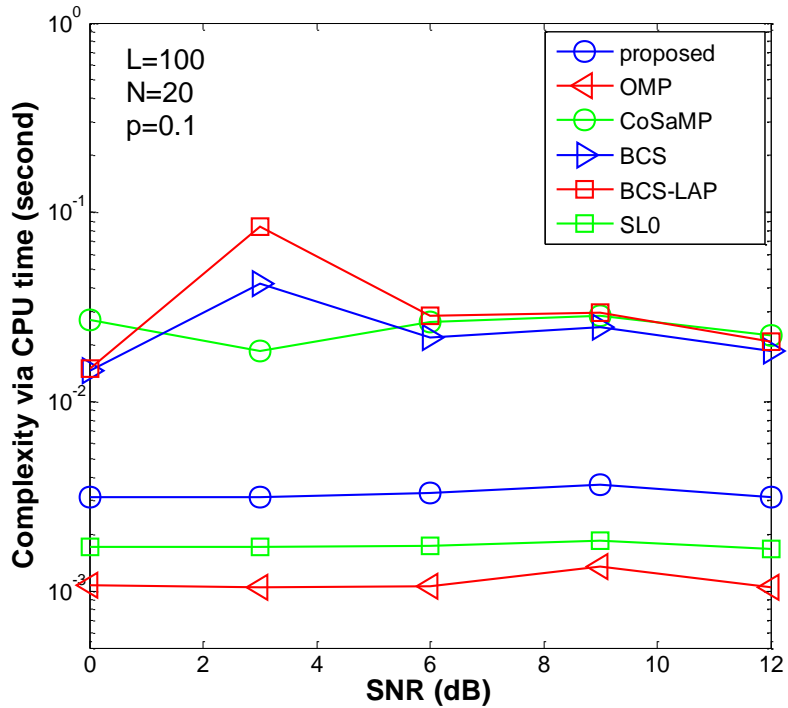

Fig. 13. Computational complexity comparison via CPU time when $p_1 = 0.1$ and $N = 20$.

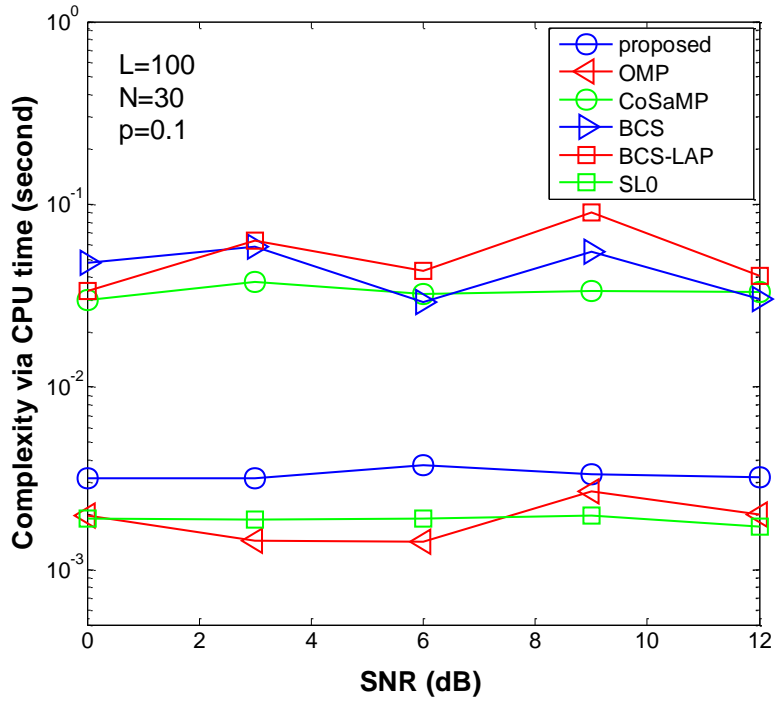

Fig. 14. Computational complexity comparison via CPU time when $p_1 = 0.1$ and $N = 30$.

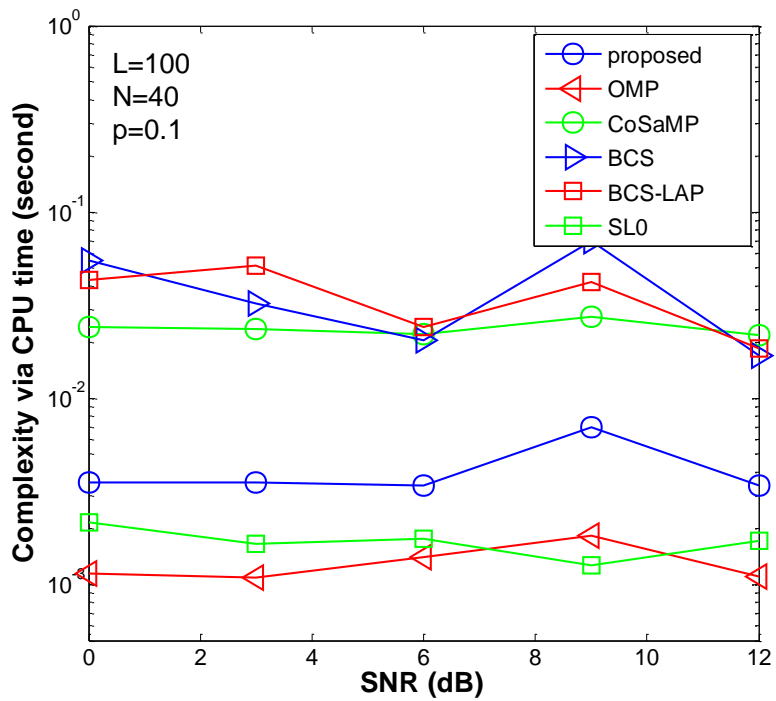

Fig. 15. Computational complexity comparison via CPU time when $p_1 = 0.1$ and $N = 40$.

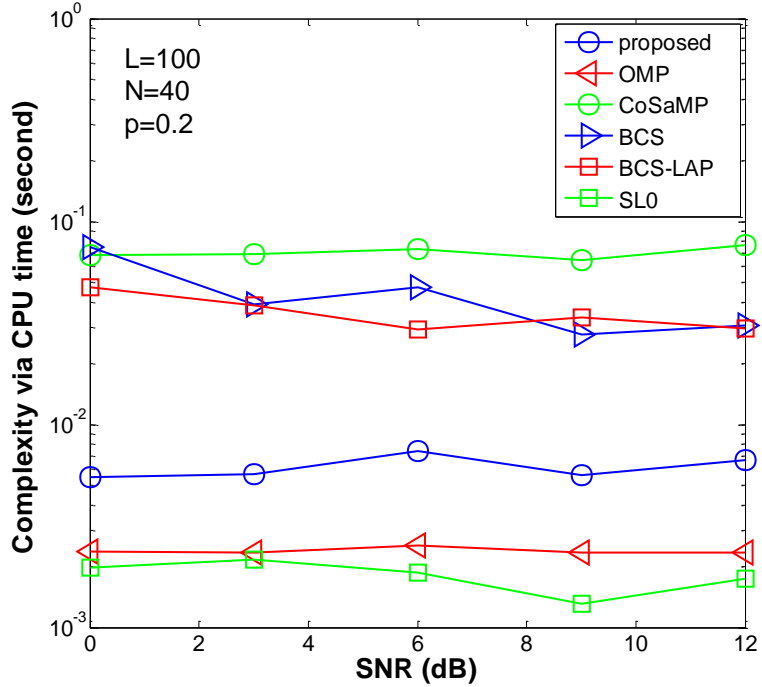

Fig. 16. Computational complexity comparison via CPU time when $p_1 = 0.2$ and $N = 40$.

## 5. Conclusion

Traditional sparse channel estimation methods are vulnerable to noise and column coherence interference in training matrix. Their primary aim is tried to exploit sparse structure information without a report of posterior channel uncertainty. To improve the estimation performance, fast Bayesian matching pursuit algorithm with application to sparse channel estimation has not only exploited the channel sparsity but also mitigated the unexpected inferences in training matrix. In addition, the propose method has revealed potential ambiguity among multiple channel estimators that are ambiguous due to observation noise or correlation among columns in the training signal. Computer simulation results have showed that propose method improved the estimation performance with comparable computational complexity when comparing with traditional methods.

## Competing interests

The authors declare that there is no conflict of interests regarding the publication of this paper.

Authors of the paper do not have a direct financial relation that might lead to a conflict of interests for any of the authors.